\documentclass[preprint,review,12pt,sort&compress]{elsarticle}

\usepackage{epsfig}

\usepackage{amssymb,amsmath}

\usepackage{lineno}

\usepackage[british]{babel}
\usepackage{array}

\begin{document}
\begin{frontmatter}



\title{Verification, Validation and Testing of Kinetic Mechanisms of Hydrogen Combustion in Fluid Dynamic Computations}


    \author{Victor P. Zhukov}
    \address{Institute of Space Propulsion, German Aerospace Center (DLR), Lampoldshausen, 74239 Hardthausen, Germany. Ph.: +49-(0)6298-28-633, Fax: +49-(0)6298-28-458}
    \ead{vpzhukov@gmail.com}

\begin{abstract}
A one-step, a two-step, an abridged, a skeletal and four detailed kinetic schemes of hydrogen oxidation have been tested. A new skeletal kinetic scheme of hydrogen oxidation has been developed. The CFD calculations were carried out using ANSYS CFX software. Ignition delay times and speeds of flames were derived from the computational results. The computational data obtained using ANSYS CFX and CHEMKIN, and experimental data were compared. The precision, reliability, and range of validity of the kinetic sche\-mes in CFD simulations were estimated. The impact of kinetic sche\-me on the results of computations was discussed. The relationship between grid spacing, timestep, accuracy, and computational cost were analyzed.

\end{abstract}


\end{frontmatter}

\linenumbers

\section{Introduction}

In the last decade commercial software packages: ANSYS CFX \cite{ANSYSCFX}, Fluent \cite{Fluent}, Star--CD \cite{Star-CD}, etc. are widely used as a tool for solving Computational Fluid Dynamics (CFD) problems. According to web site top500.org, computational capabilities are increasing by a factor of 100 every eight years. The explosive growth of computational power allows to carry out CFD simulations using more and more complicated physical models on small clusters of computers or even desktop computers (without using high-power expensive computers). The modern CFD simulations can be multicomponent, multiphase and multidomain. Heat, mass and radiation transfer as well as chemical processes can be taken into account in calculations. The increased amount of model assumptions and parameters is an significant source of errors and faults.

\begin{figure}[hp]
	\centering
    \includegraphics[width=1.0\columnwidth]{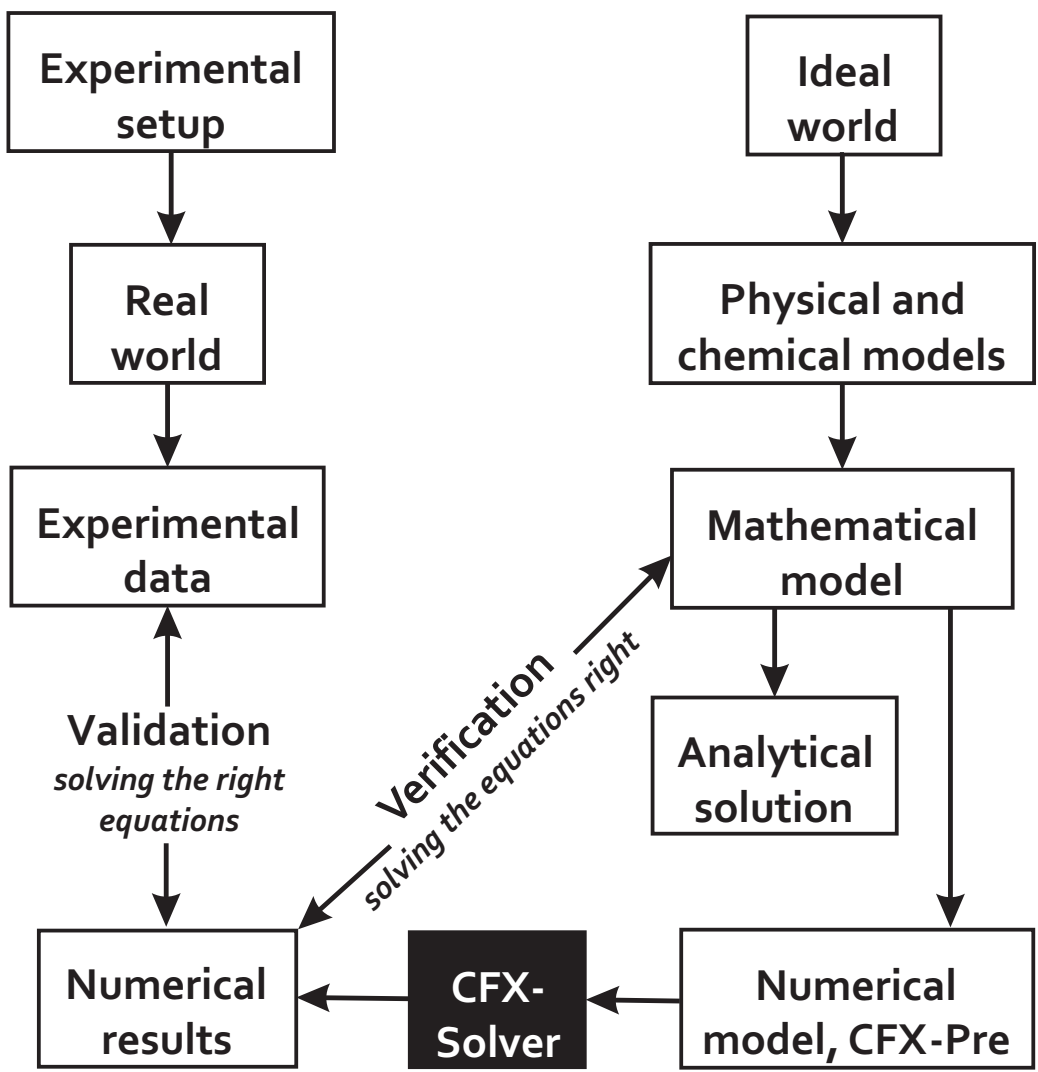}
	\caption{Logic scheme of validation and verification.}
	\label{v&v}
\end{figure}

The processes of verification and validation are very important in CFD \cite{v&v}. They are ground steps in obtaining a numerical solution (Fig. \ref{v&v}). The validation should be done prior to the obtaining of the desired numerical results while the verification should be done prior the validation. Normally, the whole numerical model, which includes equations of fluid dynamics, equation of state and the model of turbulence, is already verified by the developer of the CFD code and the user should verify only its own user defined models. Our ultimate aim is the modeling of the flow in a rocket combustion chamber. In this case we cannot rely on the predefined numerical model, but should use the models which takes into account the specifics of this complicated problem. Here we are focusing on the usage of the chemical kinetic models of hydrogen combustion. In most cases the assumption of thin flame (infinitely fast chemical reactions) gives reliable results, so there is no actual need to use the detailed kinetic mechanisms in CFD simulations. However, the assumption of thin flame is not completely satisfied in rocket combustion chamber where the turbulence is very high. By this reason the model of the chemical kinetics should be used for the modeling of the combustion in rocket engine, but before the model should be verified and validated.

In our case the verification can be done through the comparison with Chemkin \cite{CHEMKIN} which solves a system of kinetic equations. This gives us a chance to find and eliminate misprints and to prove that numerical parameters, for example time-step and grid, do not determine the solution. The next step should be the validation. After entering into a CFD code a chemical kinetic model became a part of large physical--chemical numerical model. Generally, kinetic mechanisms are already validated extensively by their authors, but after the implementation of the chemical kinetic model the CFD numerical model needs the validation. Of course chemical reactions drive combustion, but indeed combustion processes depend on heat and mass transfer too. Although turbulence model, equations of state, transport coefficients, chemical kinetic mechanism can be validated separately, the resulting physical--chemical model needs the final validation as a whole.
 
Probably the first example of the verification and the validation of hydrogen reaction mechanism  in CFD simulations  is the work by Mani {\it et al.} \cite{Mani}. A supersonic flow in a constant-area channel was simulated. The employed kinetic scheme consisted of 8 reactions without the kinetics of the peroxides. The supersonic combustion of a hydrogen--air mixture at a high temperature was modeled. The simple kinetic scheme reproduced experimental data properly, what is not surprising when the initial temperature is about 1400 K.

In 1988 Jachimowski \cite{Jachimowski} modeled scramjet combustion using of a hydro\-gen--air reaction mechanism. He carried out the simulations of hydrogen combustion at parameters related to flight Mach number 8, 16 and 25. The hydrogen--air reaction mechanism consisted of 33 reactions, where the detailed hydrogen--oxygen kinetics consisted of 20 reactions. In his study it was shown that chemical kinetics of HO$_2$ is important at all the studied Mach numbers. Later Eklund and Stouffer \cite{Eklund} carried out the 3-D CFD simulation of a flow in a supersonic combustor. They tested two kinetic models: the detailed model by Jachimowski and the model abridged from the detailed one. This model was obtained by cutting the kinetics of HO$_2$ and H$_2$O$_2$. The new model consists of only 7 reactions within 6 species plus bath gas N$_2$. This abridged model is very extensively cited and used due to the low required computational power in the comparison with detailed kinetic models.

Kumaran and Babu \cite{Kumaran} studied the effect of chemical kinetic models on CFD calculations. They modeled a compressible, turbulent, reacting flow, which simulates the supersonic combustion of hydrogen in the jet engine of hypersonic projectile. The idea of their work is to compare the results obtained using the detailed kinetic model with the previous results obtained using a single step kinetics. In their past study they attributed the difference between the numerical results and the experimental data to the inadequacy of the one-equation turbulence model and the one-step chemistry. The simulations with the use of the detailed kinetic mechanism predicts higher combustion efficiency than the calculations using the one-step model. The comparison of wall static pressure from  the experiments and the numerical simulations showed that the detailed and the single step chemistry give similar results and predict well the positions of pressure peaks, but they fail to predict the values of pressure peaks. All the numerical simulations were carried out using software package Fluent \cite{Fluent}.

The use of a detailed mechanism should give a more precise estimation of the main thermodynamic parameters: temperature and pressure, and should provide the distribution of the intermediates: OH, H$_2$O$_2$, etc. Generally detailed kinetic model provides much more information about combustion processes, but the use of it costs an additional CPU power.

In the considered above work Kumaran and Babu used the kinetic mechanism of Stahl and Warnatz \cite{WarnatzH2} published in 1985 as the reference detailed mechanism. This mechanism became a little bit old after publication of works \cite{HO2,OH} in 2002. In these works the refined data on the rate constant of reaction R9 \cite{HO2}
\begin{equation*}
\rm H + O_2 + M \longrightarrow \rm HO_2 + M \eqno \textrm{R9}	
\end{equation*}
and the enthalpy of formation of OH \cite{OH} were reported. Since radicals OH and HO$_2$ play the essential role in hydrogen oxidation, all kinetic mechanisms developed before 2002 should be revised. In the current work the several detailed hydrogen kinetic mechanisms are tested. One is ``old'' \cite{Gutheil}, while three others \cite{Konnov2008,Henri,DryerNew} were released after 2002. Besides the outdated thermodynamic data and the outdated rate constants ``old'' mechanism \cite{Gutheil} has reaction:
\begin{equation*}
\rm H_2 + O_2 \longrightarrow \rm OH + OH. \eqno \textrm{RX}	
\end{equation*}
The usage of this reaction became marginal nowadays, for example: it is not included into modern mechanisms \cite{Konnov2008}-\cite{Henri} considered below. The ``old'' mechanism has been tested in order to see the difference from the updated mechanisms.

Konnov reported recently about ``remaining uncertainties in the kinetic mechanism of hydrogen combustion'' in work \cite{Konnov2008}. He studied the detailed hydrogen combustion mechanism. Konnov found two groups of the uncertainties. The first group is associated with the set of the chemical reactions in the hydrogen--oxygen system. Not all of the possible reactions are included in the kinetic mechanisms, and the set of reactions varies from one mechanism to another. Thereby there is no one conventional set of reactions, which describes combustion of hydrogen comprehensively, and this problem is still open. The second type of uncertainties relate to the uncertainties in the rate constants of the employed reactions. Some of them are not well defined or the experimental data on them are controversial. It should be noted that developed by Konnov hydrogen kinetic mechanism \cite{Konnov2000} has the slightly different set of reactions from others extensively cited mechanisms from Princeton University \cite{Dryer}, National University of Ireland, Galway \cite{Henri}, etc. 

Shatalov  et al. \cite{Shatalov} carried out the analysis of several detailed kinetic mechanisms of hydrogen combustion: the mechanism by Konnov \cite{Konnov2008} and mechanisms by other authors. Shatalov et al. \cite{Shatalov} noted that the use of reaction RX does not have a sense, because another parallel channel
\begin{equation*}
\rm H_2 + O_2 \to \rm H + HO_2 \eqno \textrm{R-10}	
\end{equation*}
has the significantly greater rate constant (in 50 and more times). On the other hand the use of reaction RX helps a lot to fit experimental data. They pointed out also that below 1100 K reaction R9 effects on the ignition delay time significantly.

There are two very recent hydrogen reaction mechanisms from Princeton and Stanford universities \cite{DryerNew,Hong}. The both are validated against the latest experimental data. For our applications (rocket combustion) the experimental data of Burke and coworkers \cite{Burke} (the same team as \cite{DryerNew}) is the most interesting. This experimental data represents the measurements of the burning velocities in H$_2$/O$_2$/He mixtures at pressures 1--25 atm. These measurements are the only available data for hydrogen at high pressure. The comparisons of these recent mechanisms with the experimental data showed that the mechanism by Burke {\it et al.} \cite{DryerNew} has a better agreement with experimental data at high pressure than mechanism \cite{Hong}. By this reason only mechanism by Burke {\it et al.} \cite{DryerNew} has been chosen for the tests in this work.

A study similar to the current work was carried out by Gerlinger {\itshape et al.} \cite{Gerlinger}. The colleagues studied several hydrogen/air reaction mecha\-nisms including multi-step schemes \cite{Henri,Jachimowski,Eklund} and one-step mechanism by Marinov {\it et al.} \cite{Westbrook}. The study is focused in the application of reaction mechanisms in the simulation of supersonic combustion. The mechanisms were validated against ignition delay times. In the validation all mechanism showed similar results excluding one-step mechanism~\cite{Westbrook}, which missed a non-Arrenius behavior of the experimental data. The authors simulated supersonic combustion with the different mechanism and compared the results with an experimental data. They studied the influence of timestep and numerical grid as well. The numerical results showed the sensitivity to the timestep and the grid density. Finally the authors conclude that one-step mechanism \cite{Westbrook} is not appropriate, while mechanism by O'Conaire {\it et al.} \cite{Henri} is more precise than other multi-step mechanisms.

While the examples of successful verification, validation and application of hydrogen reaction mechanisms in CFD simulation of supersonic combustion ramjet exist, the problem of the CFD simulation of hydrogen combustion in rocket engine is not closed. Scramjet is a specific case and the results obtained for supersonic combustion cannot be extended over the case of rocket engine. Combustion in rocket engine has its own characteristic features: high pressures (50--250 atm), the wide span of temperatures from 100 K to 3500 K, the absence of dilutant (nitrogen). In the case of scramjet the verification and validation can be done by simulating a supersonic combustion directly what is not possible in the case of combustion in rocket engine yet. In recent work \cite{Oefelein} the group of researchers from five research centers made the CFD simulations of a flow in a combustion chamber. The each participant of the project modeled the same object using own methodology. It was the sub-scale rocket engine with 1.5 inch inner diameter, with one co-axial injector. The combustion chamber had an axial symmetry, which allowed to carry out the comparison of 2D and 3D modeling. The authors compared steady Reynolds--Average Navier--Stockes (RANS), unsteady Reynolds--Average Navier--Stockes (URANS) and three different Large Eddy Simulation (LES) models with the experiment. The comparison showed that all approaches give the noticeably different results and the only in one case (LES --- stochastic reconstruction model) the obtained results were comparable with the experimental data. Indeed the most precise modeling results were obtained with the finest mesh of $255\cdot10^6$ cells and using the highest computational power of 2 million cumulative CPU hours. However at the current moment it is not totally clear how the initial model assumptions affected the accuracy of the final results, and what assumption or parameter impaired the other models. Such comparison is very important from the practical point of view because the computational cost and precision vary strongly from one numerical model to another.

The performance of the chemical kinetic models of hydrogen oxidation in CFD simulations has been estimated in the current work. The aims of the work are the ranking of the selected hydrogen kinetic models and the development of the verification, validation and ranking procedures. In the current work the performance of the kinetic models are assessed using the experimental data on hydrogen ignition \cite{ignition} and hydrogen flame \cite{flame}. The CFD simulations are carried out using complex physical models. In such conditions the all-round verification is simply obligatory for the CFD modeling, while there is no conventional way to verify combustion models as a part of the whole physical-chemical model as well as no conventional way to verify and validate physical model itself so far. The simulations have a secondary aim to estimate the validity region within the space of the computational parameters: mesh, computational scheme, time step. The precision (the difference between calculations and experiments) and the computational cost (required CPU time) were estimated on the each test case. Global reaction model \cite{Westbrook}, two-step scheme \cite{2step}, abridged Jachimowski's model~\cite{Eklund}, a new skeletal mechanism, four detailed hydrogen mechanisms \cite{Gutheil,Konnov2008,Henri,DryerNew} have been tested. Thus the results of the work should show what chemical kinetic scheme should be used, at which parameters  the scheme should be used, how much computing power it is necessary to have for the fulfilment of a task. The work is the first step before the CFD simulations of the experiments carried out at our test facility \cite{Oschwald,Suslov}.

\section{Skeletal kinetic model}
\label{sec:KineticModel}
In our case as well as in other CFD simulations the numerical and the physical models have a lot of parameters and need the debugging before getting the final solution. Parameters, which are not connected with the problem directly, for example timestep, can seriously obstruct the obtaining of a solution. Detailed kinetic mechanisms make CFD models too heavy for the debugging. On the other side global reaction models do not include the minor species and do not allow to do the all-round debugging (validation and verification of the properties of HO$_2$, H$_2$O$_2$, etc.).

In this work a skeletal kinetic scheme has been developed, which has the same set of species as detailed hydrogen mechanisms, but the reduced set of reactions. This light scheme sped up the formulation of the computational problem. The problem definition requires to perform the certain set of the calculations. Different meshes and the models of diffusion and thermal conductivity were tried before getting the final results. The light skeletal scheme reduced the amount of the expended CPU hours at the preliminary stage. The new scheme fills the gap between abridged Jachimowski's model \cite{Eklund}, which has 7 reactions and 6 species, and detailed hydrogen mechanisms \cite{Gutheil,Konnov2000,Henri} (19--21 reactions, 8 species and bath gases) as well as it allows to separate the influence of the amount of reactions and species.

The new scheme was developed from the skeletal model by Kreutz and Law~\cite{9step}. Their skeletal kinetic model has 9 unidirectional reactions and 8 species. Considering H$_2$/O$_2$ system it may assume that the set of 9 species: H$_2$, O$_2$, H$_2$O, H, O, OH, HO$_2$, H$_2$O$_2$ and bath gas is complete and other species (O$_3$, OH$^-$, OH$^*$(A), etc.) can play role only in marginal cases. It is necessary to note that at high pressures, what is the case of rocket combustion chambers, the chain branching proceeds via the formation of HO$_2$, H$_2$O$_2$ radicals due to the high rates of the recombination processes \cite{Zhukov}. For example reaction
\begin{equation*}
\rm H + O_2 \longrightarrow \rm OH + O, \eqno \textrm{R1}	
\end{equation*}
which is the most important in atmospheric hydrogen flames, is suppressed by reaction R9 at pressures above 50~bar. The model by Kreutz and Law \cite{9step} has a 5~times smaller set of reactions as detailed hydrogen mechanisms and can adequately predict the ignition delay times and the ignition limits. On the other hand the scheme consists of the irreversible reactions, which means that the concentrations of species never reach the equilibrium state. The afterburning processes are omitted, which is not important during ignition, but leads the mispredictions of species profiles. By these reason the reaction set was extended by 6 reactions from detailed hydrogen model~\cite{Henri}. The reaction of the quadratic recombination of HO$_2$ radicals  
\begin{equation*}
\rm HO_2 + HO_2 \longrightarrow \rm H_2O_2 + O_2 \label{R14}	
\end{equation*}
was substituted by reactions R11 and R13, see Tables \ref{tab:model}, \ref{tab:eff_factors}. Such extension increases the computational weight of the model, but it increases the adequacy of the model as well. The new added reactions involve the processes of radical recombinations, which are important in a post flame zone.

\section{Calculations}
\label{sec:Calculations}
The CFD calculations have been done with the use of the ANSYS CFX 11 solver~\cite{ANSYSCFX}, which utilizes the Finite Volume Element Method (FVEM). The meshes have been created using ANSYS ICEM software. The choice of the software is given by an adherence to the compatibility of the computer data and the design documentation.

Two types of tests (simulations) have been done in the work. The first test case is a quasi 0-dimensional simulation of hydrogen ignition to verify and validate the models against the experimental data on ignition delay times \cite{ignition}. The second test case is an 1-dimensional simulation of hydrogen flame propagation to test the models against the data on the speeds of laminar flame \cite{flame}.

An ignition in a perfect adiabatic constant volume reactor has been modeled as the quasi 0D problem. By the formulation the problem is dimensionless, but by the settings of calculations it is 3D. The computational domain represents eighth part of the 1 mm sphere with rigid adiabatic walls. The mesh consists of 21~nodes and 38 tetrahedron elements. At the initial moment the whole domain is filled with a stoichiometric hydrogen--air (0.79N$_2$+0.21O$_2$) mixture at pressure of 1 atm and temperature in the range of 900--1400 K. The problem has been solved as a transient task, i.e. the time evolution of gas conditions has  been sought. The laminar model (unsteady Navier--Stokes equations) was employed, but also one series of simulations were carried out using the \emph{k-$\epsilon$} turbulence model. The comparison of results shows that both models (the laminar and the \emph{k-$\epsilon$} model) give the same results in this task. The task was imposed in such way that the results should be independent on the choice of turbulence model. Indeed the stagnant homogeneous gas mixture is surrounded with the adiabatic rigid walls, so the gas inside the sphere should be stagnant all time. The object of these calculations is the estimation of the ignition delay times and the comparison of the calculated delay times with the data from shock tube experiments \cite{ignition}, see Fig. \ref{delays}. In the calculations the ignition delay times were defined as the time of a temperature increase up to 500 K relative to the initial temperature.

During the 1D tests a freely propagating hydrogen flame has been modeled. The computational domain consists of 1604 nodes and 400 rectangular prism elements. All elements are placed along one axis so that the thickness of the domain equals to one element in two other coordinate axes. The mesh spacing equals to 5~$\mu$m in the direction of the flame propagation. The separate study of the influence of the grid spacing was carried out where the spacing was varied from 0.2~$\mu$m to 200~$\mu$m. The domain represents the rectangular with symmetry boundary conditions on the side walls. The domain has one inlet and one outlet (on the side opposite to inlet). At the outlet static pressure is specified and equals to 1 atm. At the inlet a hydrogen--air (0.79N$_2$+0.21O$_2$) mixture at 298 K and 1 atm flows inside the domain. The velocity of the mixture is specified at the inlet in the range of 0.5--3.5~m/s so that the velocity of the flame front reaches a small value in the laboratory system of coordinates. The mixture composition was varied from equivalence ratio of ER = 0.5 to ER = 4.5. The simulations were run as a transient task. A stationary burning velocity was sought. The laminar model was employed, also one series of tests were carried out using the \emph{k-$\epsilon$} turbulence model. Speed of flame depends essentially on the transport properties of gas, so the temperature dependent thermal conductivity and diffusion coefficients were used. Thermal diffusion was not taken into account. The system of the governing equations in ANSYS CFX does not assume mass fluxes caused by temperature gradients.

\section{Results and discussion}
\label{sec:Discussion}
\subsection{Verification. Comparison with CHEMKIN.} 
\label{sec:ChemkinII}

The both tasks were also solved in CHEMKIN II \cite{CHEMKIN}. The results of the simulations with the help of CHEMKIN II were used as a reference data. CHEMKIN is very widely used for solving chemical kinetic problems, where the computational problem is formulated as solving of a system of ordinary differential equations. Indeed ASYS CFX allows to specify the properties of a system by the different ways, while in CHEMKIN task is set in the one prescribed format. CHEMKIN uses the modified Arrenius form for rate coefficients:
\begin{equation*}
k = A\cdot T^{n}\cdot\exp(-E_a/RT). \label{Arrenius}
\end{equation*}
The thermodynamic functions: enthalpy, entropy and heat capacity are calculated using the NASA polynomial forms in CHEMKIN. During the calculations in ANSYS CFX the same equations were employed for the rate constants, the thermodynamic functions and the equations of states, so the comparisons of the results obtained using the different software have a sense. The results of the comparisons is depicted in Fig. \ref{chemkin0D} and Fig. \ref{chemkin1D}, where ANSYS CFX shows the agreement with CHEMKIN.

Using CHEMKIN the ignition delay times were calculated in the assumptions of constant volume and adiabatic walls. In this case the problem definitions (the sets of boundary conditions and kinetic equations) correspond to each other in ANSYS CFX and CHEMKIN. As a consequence the results of the simulations using ANSYS CFX agree fully with the computations in CHEMKIN, see Fig. \ref{chemkin0D}. Indeed it is necessary to note that CFX solves the 3-dimensional Navier--Stokes equations while CHEMKIN uses the 0-dimensional equation of energy conservation.

The next step is the modeling of freely propagating laminar flame. Zeldovich--Frank-Kamenetskii equation, which connects flame velocity and reactivity, gives us a clear view on the problem:
\begin{equation}
	u_{\textrm lam}=\sqrt{\frac{\alpha}{\tau}},\label{ZFK}
\end{equation}
where $\tau$ is the chemical time scale in reaction zone, and $\alpha$ is the coefficient of temperature conductivity, which summarizes the effect of diffusion and heat conductivity through Lewis number $Le=1$. In contrast to the previous case kinetic and transport properties have an equal importance in flame propagation.

The flame speeds were estimated using ANSYS CFX and PREMIX \cite{PREMIX}, Fig.~\ref{chemkin1D}. PREMIX is a subroutine of the CHEMKIN which computes species and temperature profiles in steady-state laminar flames. The transport properties were estimated using TRANFIT: the another part of the CHEMKIN collection. Thermal conductivity, viscosity and diffusion coefficients are estimated from the parameters of the Lennard--Jones potential and the dipole moment of species. The same temperature depended coefficients of thermal conductivity, viscosity and binary diffusion were used in PREMIX and ANSYS CFX, but diffusion fluxes in multicomponent mixture were approximated by a different way.

The specific of H$_2$--O$_2$ system is that the properties of hydrogen (the lightest gas) distinguishes strongly from the properties of other species within the system. By default ANSYS CFX estimates the transport properties in the mixture of gases by an inappropriate way for H$_2$--O$_2$ system, where the influence of the fuel on the transport properties is important. It calculates the coefficient of thermal conductivity and viscosity of gas mixture using a mass averaging, and the coefficients of diffusion are calculated from the mixture bulk viscosity. The problem becomes significant in the case of combustion in rocket engine where the mixture is not diluted by nitrogen. This problem can be resolved in ANSYS CFX using CFX Expression Language (CEL) and setting all coefficients by user as it was done in this work.

In H$_2$--O$_2$ system the diffusion coefficients vary from one component to other in $\sim \,$6 times: $D_{O_2}/D_H \approx \sqrt{(\mu _{O_2} /\mu _H)} \approx \sqrt{32}$. Even in the simplest case, where only H$_2$, O$_2$, H$_2$O are taken into account, the gas mixture can not be assume as a binary mixture or as a solution of light gas in heavy gas due to the high fractions of H$_2$ and H$_2$O and the differences in the diffusion coefficients. In the current work the diffusion coefficients are calculated by the empirical formula 
\begin{equation}
D_i=\frac{1-w_j}{\sum X_j/D_{ij}}	, \label{diffusion}
\end{equation}
where $w_i$ is the mass fraction of i-species; $X_j$ is the mole fraction of j-species; $D_{ij}$ is the binary diffusion coefficient \cite{Warnatz}. After that the diffusion coefficients of individual species are put into the equation which is responsible for the transport in CFX:
\begin{equation}
\rho_i(U_{mix}-U_i)=-\frac{D_i}{\rho_{mix}}\frac{\partial \rho}{\partial x}	, \label{transport_CFX}
\end{equation}
where $\rho_i(U_{mix}-U_i)$ is  the relative mass flux of i-species. The equation is not solved for an one constraint component (in our case nitrogen), which mass fraction is calculated from the constraint that the sum of mass fractions of all species equals to 1. PREMIX (CHEMKIN) uses a more accurate definitions of the diffusion and  the thermal conductivity in gas mixture, and takes into account the thermal diffusion of H and H$_2$. There are two options: ``mixture-averaged properties'' and ``multicomponent properties'' in PREMIX.  ``Mixture-averaged'' option, which was used here, employ eq. \eqref{diffusion}, but does not have a constraint species and employs an additional term --- correction velocity, which makes the net species diffusion flux equal to zero. ``Multicomponent'' formulation uses the method described by Dixon-Lewis~\cite{Dixon-Lewis}, where the coefficients are computed from the solution of a system of equations defined by the $L$ matrix.

The flame velocities obtained with the use of CFX and CHEMKIN coincide practically with each other. The difference in the results, which is small (Fig. \ref{chemkin1D}), should be associated with the distinction between the formulations of the diffusion fluxes. Coffee and Heimerl \cite{Coffee} compared various methods of approximating the transport properties of premixed laminar flames, in particular the methods which have been used in CFX and CHEMKIN. They found that the difference in flame speed is small for these methods, but the method, which is employed in CHEMKIN, is more accurate than the method with constrained species (CFX), which is inaccurate in computing the diffusion velocity for constrained species. As for the comparison with experimental data it was shown in recent work \cite{Yang} that such small overshooting around the stoichiometry, which is observed in Fig. \ref{chemkin1D}, results from the neglecting Soret effect (thermodiffusion).

In our case (laminar hydrogen--air flame at 1 atm) the propagation of the flame is supported essentially by the diffusion H$_2$ into flame zone and the diffusion of H into preflame zone. That is why it is important to estimate accurately the diffusion term. in Fig.~\ref{flamesp} we can see the effect of the transport properties on laminar flame. The maximum of the flame speed is shifted to the higher equivalent ratios where the diffusive fluxes of H and H$_2$ are higher.

\subsection{Validation and testing}
\label{sec:validation}
\begin{figure}[hp]
	\centering
    \includegraphics[width=1.0\columnwidth]{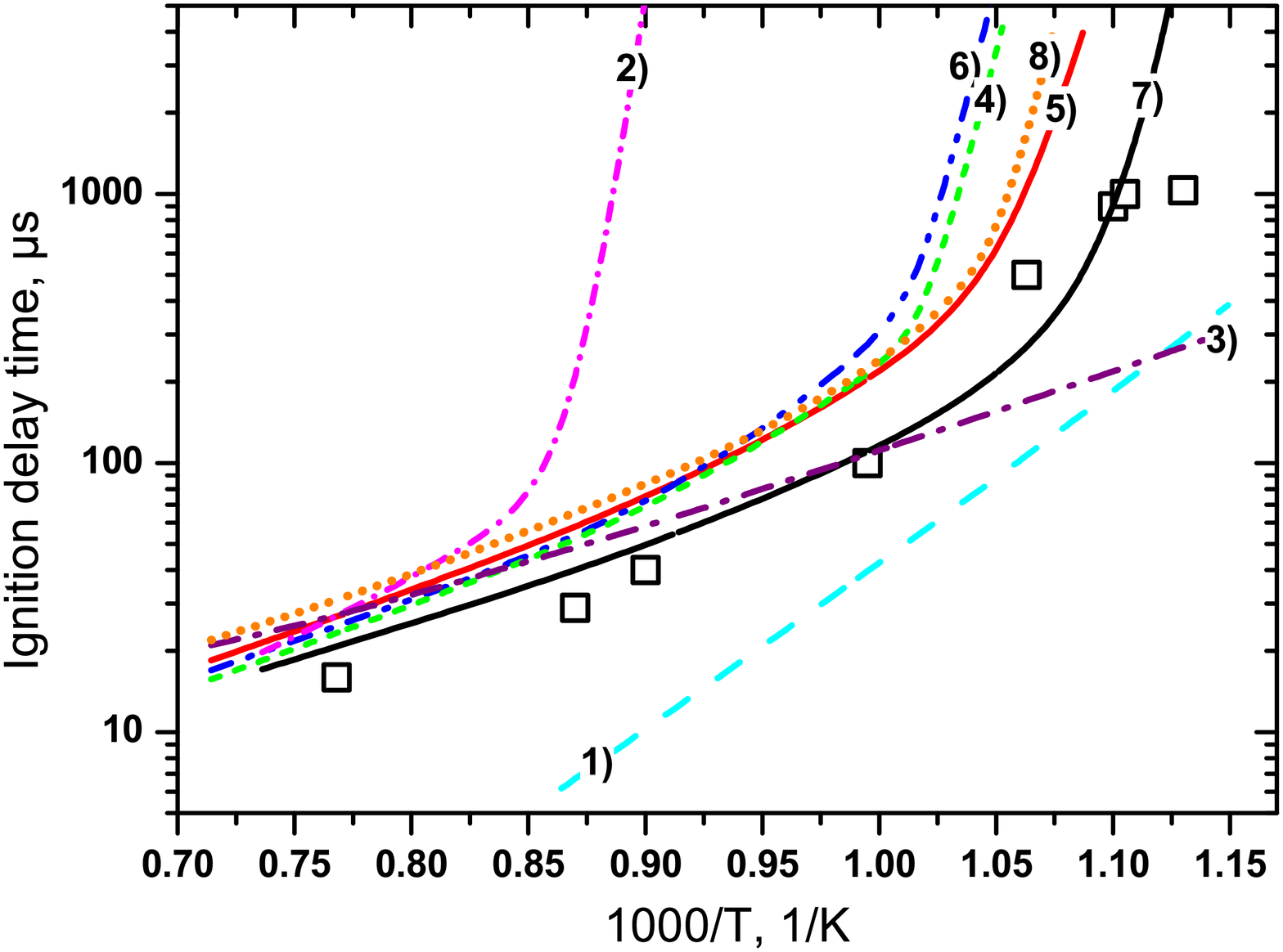}
	\caption{Ignition delay times of a stoichiometric hydrogen--air mixture at 1~atm. Squares ---  experimental data \cite{ignition}; 1) dash cyan line   --- Marinov {\it et al.} \cite{Westbrook}; 2) short dash dot magenta line ---  Lee and Kim \cite{2step}; 3) dash dot purple line --- the abridged Jachimowski's model \cite{Eklund}; 4) short dash line green  --- Zhukov (this work); 5) solid red line ---  O'Conaire {\it et al.} \cite{Henri}; 6) dash dot dot line blue ---   Gutheil {\it et al.} \cite{Gutheil}; 7) solid black line (the closest to exp. data) ---  Konnov \cite{Konnov2000}; 8) short dot orange line --- Burke {\it et al.} \cite{DryerNew}.}
	\label{delays}
\end{figure}

\begin{figure}[hp]
	\centering
    \includegraphics[width=1.0\columnwidth]{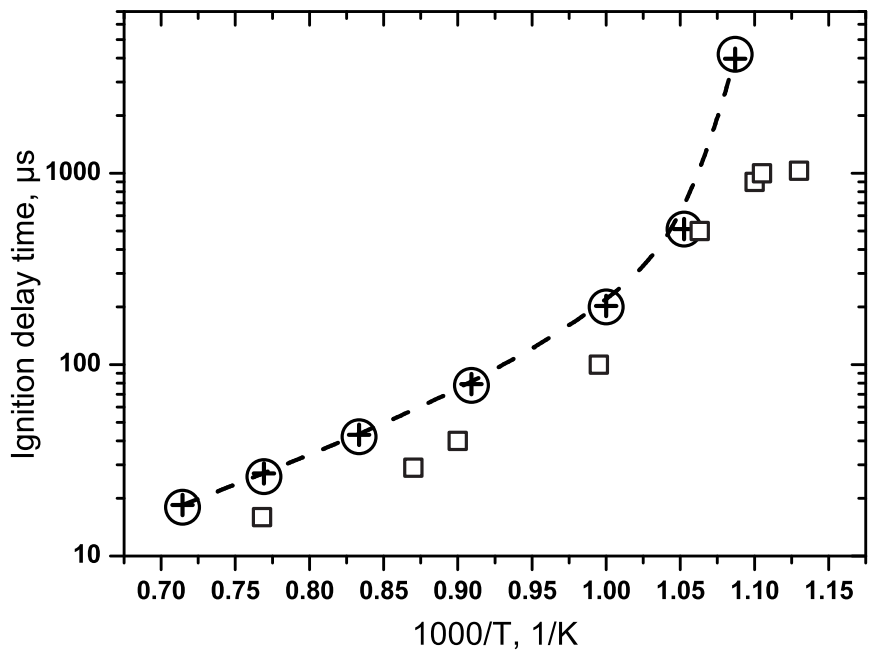}
	\caption{Comparison of the simulating data obtained using ANSYS CFX and CHEMKIN. Squares  ---  the experimental ignition delay times of a stoichiometric hydrogen--air mixture at 1~atm~\cite{ignition}; the kinetic model by O'Conaire {\it et al.}  \cite{Henri}: dash line (B-spline) and crosses --- ANSYS CFX, big circles --- CHEMKIN.}
	\label{chemkin0D}
\end{figure}

\begin{figure}[hp]
	\centering
    \includegraphics[width=1.0\columnwidth]{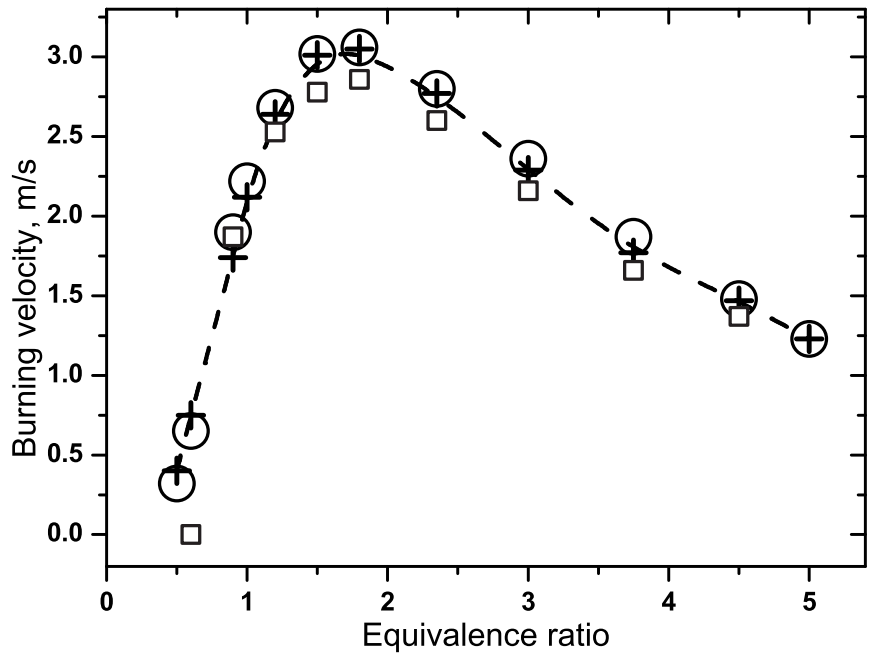}
	\caption{Comparison of the simulating data obtained using ANSYS CFX and CHEMKIN. Squares --- the experimental burning velocities of a hydrogen--air mixture at 1~atm  and 298~K~\cite{flame}; the kinetic model by Konnov \cite{Konnov2000}: dash line (B-spline) and crosses --- ANSYS CFX, big circles --- CHEMKIN.}
	\label{chemkin1D}
\end{figure}

Let us consider the results of the first ``0D'' test case, which is depicted in Fig. \ref{delays}. The detailed models \cite{Gutheil,Konnov2000,Henri,DryerNew} agree with experimental data well, while non-detailed kinetic models \cite{Westbrook,2step}, abridged Jachimowski's model \cite{Eklund} and the new skeletal model have the agreement with experimental data only in the limited range. The kinetic model by Konnov \cite{Konnov2008} agrees with experimental data better than other models. In Fig. \ref{delays} it is possible to see the transition from high--temperature kinetics to low--temperature kinetics around 950 K. Generally the models show the common trend: more details --- higher accuracy. This conclusion is supported by the results of the 1D test case too. It is possible to conclude from the results that one or two reactions are not enough to describe the ignition of hydrogen. Probably the sophistication of abridged Jachimowski's model (7 reactions and H, O, OH as intermediates) is a reasonable minimum for the modeling of hydrogen combustion in the high temperature region ($T>1000 \, \mathrm K$). For the modeling in a wide temperature range the formation of H$_2$O$_2$ and HO$_2$ should be taken into account. Of course it would be very surprising to see that reduced or global mechanisms can describe ignition in a wide range of parameters. They are deduced from detailed mechanisms by neglecting the marginal processes, so they can not describe the behavior near margins. In most cases the oxidation of hydrogen proceeds via the formation of HO$_2$ radical, which is not included in the reduced mechanisms. In the new mechanism the kinetics of HO$_2$ is not comprehensive too.

It is necessary to note that shock tube is not a reactor with adiabatic solid walls. Due to the boundary layer effect the temperature behind reflected shock wave slowly increases with time. The discrepancy between the ideal assumptions and reality arise at large residence times, for the most of the shock tubes after 1 ms. In Fig. \ref{delays}, where the ignition delay times are calculated using the boundary conditions of adiabatic solid walls, the detailed models have a ``wrong'' trend at low temperatures (large residence times). At large residence time it is necessary to take into account the real conditions behind reflected shock wave as it was done in \cite{Hong}. In this case the actual agreement between experiments and detailed models \cite{Konnov2000,Henri,DryerNew} will better than on the graph.

The performed simulations give more information about the evolution of the system than simply ignition delay times. The ``classical'' behavior was observed without anything unusual in the all ``0D'' tests (by this reason it is not included in the article). All the kinetic models predict similar temperature (or pressure) time-resolved profiles, which have the induction period, the following temperature (or pressure) rise, which ends with asymptotic behavior. The gas temperature (or pressure) of the combustion products is predicted correctly by the all kinetic models. 

\begin{figure}[hp]
	\centering
    \includegraphics[width=1.0\columnwidth,draft=false]{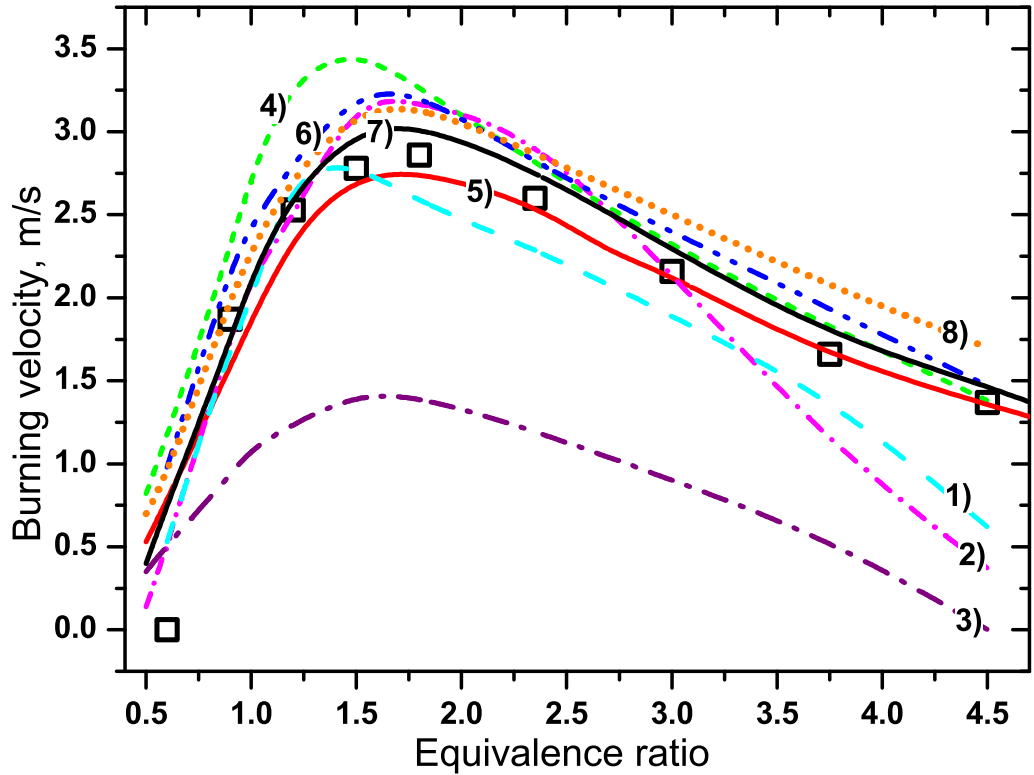}
	\caption{Burning velocities of a hydrogen--air mixture at 1~atm. Squares experimental data \cite{flame}; 1) dash cyan line --- Marinov {\it et al.} \cite{Westbrook}; 2) short dash dot magenta line ---  Lee and Kim \cite{2step}; 3) dash dot purple line --- the abridged Jachimowski's model \cite{Eklund}; 4) short dash line green  --- Zhukov (this work); 5) solid red line ---  O'Conaire {\it et al.} \cite{Henri}; 6) dash dot dot line blue ---   Gutheil {\it et al.} \cite{Gutheil}; 7) solid black line (the closest to exp. data) ---  Konnov \cite{Konnov2000}; 8) short dot orange line --- Burke {\it et al.} \cite{DryerNew}.}
	\label{flamesp}
\end{figure}

In the 1D test case the agreement of the simulating data with the experimental data is better in sum than in the ``ignition'' case, see Fig. \ref{flamesp}. Practically the all models agree with the experimental data.  The other distinctive feature of the obtained results is the bad agreement of abridged Jachimowski's model \cite{Eklund} and the good agreement of one-step model \cite{Westbrook}. The results of 1D simulations can be interpret in terms of eq. \eqref{ZFK}. The system has practically the same physical properties in the all 1D simulations. These allow us to conclude that
\begin{equation}
	\frac{u_1}{u_2}=\sqrt{\frac{\tau_2}{\tau_1}},\label{uratio}
\end{equation}
where indexes \emph{1} and \emph{2} designate the attribute to different kinetic models. The ignition delay times should be taken at flame temperature. In our case the flame temperature amounts $\sim$2000~K. Abridged Jachimowski's model \cite{Eklund} has the lowest effective activation energy among the models (see Fig.~\ref{delays}) and predicts the significantly larger $\tau$ at high temperatures. As for one-step model \cite{Westbrook}, which predicts the shortest $\tau$ at high temperatures, it does not include atomic hydrogen. It means that the assumptions, which lead us to eq.~\eqref{uratio}, are not correct for this model. In terms of eq.~\eqref{ZFK} the neglecting of the diffusion of hydrogen atom should lead to the smaller value of $\alpha$ and to the essentially less flame velocity, but it is compensated in this model by the small ignition delay time.

In Figures \ref{delays} andb \ref{flamesp} we can see the difference between the results of the simulations and the experiments. While in the case of the global or reduced kinetic models the discrepancy can be attributed to the weakness of models, detailed reaction mechanisms \cite{Gutheil,Konnov2000,Henri} represent the state-of-the-art view on hydrogen kinetics. The agreement of these models with experimental data and the role of intermediates (H, O, OH, HO$_2$ and H$_2$O$_2$) were discussed in details in original articles \cite{Gutheil,Konnov2000,Henri}. The oxidation of hydrogen proceeds via the formation of highly active intermediates. The experimental study of the kinetics of the intermediates of H$_2$/O$_2$ system has some difficulties at temperatures below 900~K where in experiments it is necessary to keep constant conditions during long residence times. Thus even a detailed kinetic scheme can fail near ignition and flammability limits.

\begin{figure}[hp]
	\centering
    \includegraphics[width=1.0\columnwidth,draft=false]{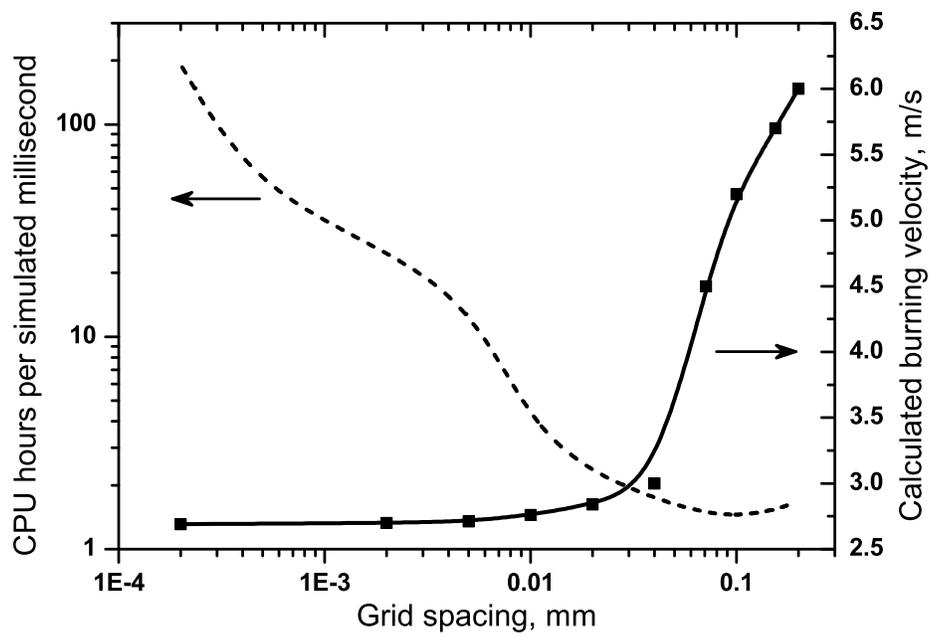}
	\caption{Simulating results and computational cost as the function of grid spacing. A hydrogen--air mixture at 1 atm, kinetic model by Zhukov (this work).}
	\label{fig5}
\end{figure}

Numerical parameters such as time step, grid spacing, type of difference scheme, etc. should not determine the results of modeling. The proper values of time step and grid spacing should correspond via the coefficients of physical model to physical time and space scales. On practical ground the upper limits of time step and grid spacing are more important, because computational cost is generally inversely proportional to timestep and the amount of nodes (for the employed grid the amount of nodes is reciprocally proportional to the grid spacing), see Fig. \ref{fig5}. The employed values of time step and grid spacing are normally close to the upper limit. Generally chemical processes can not be faster than several fractions of microsecond, and hydrodynamic processes can not take place on a scale smaller than a micrometer. Thus these values can be set as a reasonable lower limit for the time step and the grid spacing. The upper limit is quite specific to the details of a task. It is necessary to estimate the maximum time and mesh steps in each case separately. At a too high time step the solution diverges. Numerical noise and residuals can be used as the measure of the proximity to the upper limit of time step. In this work the adaptive time step has been used and it has been defined by a residual. The time step was decreased or increased until the value of the residuals reached the desired level. To estimate the upper limit of mesh step several simulations were carried out with different spacing, see Fig. \ref{fig5}. On the plot we can see a plateau for the cell size below 5~$\mu$m. The upper limit, which is located near  10~$\mu$m, is related to the flame thickness. In flame front the concentration of hydrogen increases in 2--4~times each 10~$\mu$m. Probably the maximum grid spacing is universal for all hydrogen--air flames at 1 atm and is defined by the transport properties of the system while the maximum time step is individual for each kinetic model. As we can see later, maximum time step is related to the stiffness of kinetic scheme.

Grid spacing, the physical dimensions of computational domain and computational cost are connected with each other. At these conditions grid spacing can limit the applicability of kinetic model. In Table \ref{tab:performance} the computational costs, which are required for the simulation of the evolution of the system during 1~ms on 1~CPU (Pentium 4) at 2~GHz, are presented for the all tested models. In the identical the conditions computational cost varies by orders from one model to another. It is impossible from the data of Table \ref{tab:performance} to see any direct connection of the computational cost with the number of reactions and species. However there is a trend: detailed kinetic models require a much more computational power than reduced models. Thus the high computational cost limits the application of detailed kinetic mechanisms in CFD calculations seriously. The grid spacing has the limit near 40~$\mu$m after which the simulations give the absolutely unrealistic results. Detailed models can be employed only in a special tasks with the computational domains of small sizes due to the high computational cost and the small grid spacing.

The computational cost increases strongly from the global reaction models to the detailed kinetic mechanisms, but the number of chemical equations and species does not completely determine computation cost. Another essential parameter is the stiffness of kinetic model. Stiffness is the embedded parameter of each kinetic model. It determines the maximum timestep during calculations. It arises due to differences in time\-scales for different chemical reactions. To solve kinetic equations it necessary to integrate equations using timestep related the smallest timescale over the time interval related to the largest timescale. Stiffness can be characterized  the ratio of timescales. For example, reaction R9 is in $10^{10}$ times faster than reaction
\begin{equation*}
\rm H + O_2 \longrightarrow \rm O + OH \eqno \textrm{R1}	
\end{equation*}
in preflame zone while in postflame zone it is vice versa, reaction R1 is faster than reaction R9 in 100 times. It is typical for combustion problems, when important fast reactions run on the background of slow equally important processes. Another example of the embedded stiffness is reaction
\begin{equation*}
\rm H_2O_2 + M \longrightarrow \rm OH + OH +M. \eqno \textrm{R15}
\end{equation*}
This reaction determines concentration of H$_2$O$_2$ in flame, and it changes its direction from reverse to forward after passing the flame front.

\section{Conclusions}
\label{sec:Conclusions}
The eight different kinetic models of hydrogen oxidation were verified, validated and tested in the CFD simulations what was done using ANSYS CFX 11 software. The two cases: ignition in adiabatic constant volume reactor and the propagation of free laminar flame were considered. The verification of the kinetic models was done through the comparison with the results obtained with the help of the Chemkin software. The verification allowed to eliminate misentries and to define correctly the thermodynamic, kinetic and transport properties.

The subsequent validation showed that the detailed kinetic schemes are more precise than the reduced. While it was not found any dependence between the ''speed`` of kinetic model and the number of reaction and species, the reduced kinetic scheme are faster than detailed. The simulations showed the common trend for kinetic models: more details  --- higher computational cost --- higher precision. The simulation of the ignition of hydrogen--air mixture showed that the results are sensitive to the choice of kinetic model. However in the case of the flame propagation the results are more sensitive to the model of the transport properties while the reasonable results can be achieved even with the use of global reaction mechanism.

The comparison of the simulating data with the experimental data \cite{ignition,flame} showed that detailed kinetic schemes \cite{Konnov2000,Henri,DryerNew} agree with experiments well, while the non-detailed schemes agree with the experiments only within a limited range. The kinetic model by Konnov \cite{Konnov2000} has the best agreement with the experimental data among the tested models. The application of reduced kinetic schemes of hydrogen combustion, which do not take into account chemical reactions with HO$_2$ and H$_2$O$_2$, is possible only with strong limitations.

For the debugging purposes the new skeletal kinetic scheme was developed  which represents the good compromise between computational cost and accuracy. 

The carried out study showed that computational results are affected by the parameters of physical and numerical models. A large amount of model parameters is the potential source of errors. The number of different coefficients reaches thousands in the simulation with the use of a detailed kinetic model. The parameters of the model can be verified and validated using the proposed method. 

The application of kinetic models in CFD calculations requires the considerable amount of computational power. The maximum time is limited by the stiffness of model and alters from model to model while the maximum grid spacing is more or less universal and defined by the thickness of flame front.

\section*{Acknowledgments}
\label{sec:Acknowledgments}
The author is grateful to Andreas Gernoth for introducing into ANSYS CFX. Also the author appreciate the scientific discussions with Dr. Oskar Haidn.

\newpage
\begin{table}[tp]\footnotesize
	\centering
			\caption{New skeletal mechanism.}
	\label{tab:model}
	\medskip
			\begin{tabular}{|c|c|l|c|c|c|c|}
			\hline
			\bfseries No&\bfseries Ref. No&\bfseries Reaction&\bfseries A&\bfseries n&\bfseries E$_a$&\bfseries Ref.\\
  \hline
    1&R1&$\rm H+O_2 \longrightarrow OH+O$& 1.91e+14&0.0 &16.44&\cite{9step}\\
    2&R2&$\rm H_2+O \longrightarrow H+OH$& 5.08e+4&2.67 &6.292&\cite{9step}\\
   3&R3&$\rm H_2+OH \longrightarrow H+H_2O$& 2.16e+8&1.51 &3.43&\cite{9step}\\
   4&R5&$\rm H_2+M \longleftrightarrow H+H+M$& 4.57e+19&-1.4 &105.1&\cite{Henri}\\
   5&R6&$\rm O+O+M \longleftrightarrow O_2+M$& 6.17e+15&-0.5 &0.0&\cite{Henri}\\
   6&R7&$\rm H+O+M \longrightarrow OH+M$& 4.72e+18&-1.0 &0.0&\cite{Henri}\\
   7&R8&$\rm H+OH+M \longrightarrow H_2O+M$& 4.5e+22&-2.0 &0.0&\cite{Henri}\\
   8&R9&$\rm H+O_2+M \longrightarrow HO_2+M$& 6.17e+9&-1.42 &0.0&\cite{9step}\\
   9&R10&$\rm H+HO_2\longrightarrow H_2+O_2$& 1.66e+13&0.0 &0.82&\cite{Henri}\\
   10&R-10&$\rm H_2+O_2\longrightarrow H+HO_2$& 3.68e+13&0.203 &54.46&\cite{9step}\\
   11&R11&$\rm H+HO_2\longrightarrow OH+OH$& 1.69e+14&0.0 &0.87&\cite{9step}\\
   12&R13&$\rm OH+HO_2\longrightarrow H_2O+O_2$& 2.89e+13&0.0 &-0.5&\cite{Henri}\\
   13&R15&$\rm H_2O_2+M\longrightarrow OH+OH+M$& 1.2e+17&0.0 &45.5&\cite{9step}\\
   14&R-17&$\rm H_2+HO_2\longrightarrow H+H_2O_2$& 3.42e+12&0.202 &27.12&\cite{9step}\\
      \hline
   		\end{tabular}
\begin{flushleft}
$ k = A\cdot T^{n}\cdot\exp(-E_a/RT)$;  units: mol, cm$^{3}$, K, kcal; thermodynamic data~\cite{CHEMKIN}; the reverse rate constants (R5, R6) are calculated from the forward rate constants through the equilibrium constants.\\
\end{flushleft}
		\caption{Efficiency factors for third body term.}
\label{tab:eff_factors}
	\medskip
	\begin{tabular}{|c|c|c|c|c|c|c|c|c|}
		\hline
			\bfseries Ref. No&\bfseries H&\bfseries H$_2$&\bfseries H$_2$O&\bfseries H$_2$O$_2$&\bfseries HO$_2$&\bfseries O&\bfseries O$_2$&\bfseries OH\\
  \hline
    R5&1.0&2.5& 12&1.0 &1.0&1.0&1.0&1.0\\
    R6&0.83&2.5& 12&1.0 &1.0&0.83&1.0&1.0\\
    R7&0.75&2.5& 12&1.0 &1.0&0.75&1.0&1.0\\
    R8&1.0&0.73&12&1.0 &1.0&1.0&1.0&1.0\\
    R9&1.0&2.5&12&1.0&1.0&1.0&1.0&1.0\\
    R15&1.0&2.5&12&1.0&1.0&1.0&1.0&1.0\\
                \hline
   		\end{tabular}
\end{table}
\begin{table}[tp]\footnotesize
	\centering
		\caption{Parameters of kinetic models and the computational costs. Case~1 --- test case ``ignition'', 38~cells; case 2 --- test case ``flame propagation'', 400~cells.}
	\label{tab:performance}
		\medskip
		\begin{tabular}{|m{3.1cm}|m{1.7cm}|m{2.2cm}|m{2.1cm}|m{2.1cm}|}
					
				\hline

			\bfseries \center Model&\bfseries  Number of equations$^*$&\bfseries Number of species&\bfseries Case\:1,  CPU~hours/ms&\bfseries  Case\:2, CPU~hours/ms  \\
   \hline
    Marinov {\it et al.}  \cite{Westbrook}&1&3 (H$_2$, O$_2$, H$_2$O) + N$_2$&0.18&0.14\\
   \hline
    Lee and Kim  \cite{2step}&2&4 (3 + H) + bath gas&0.43&0.20\\
    \hline
abridged Jachimowski's \cite{Eklund}&7&6 (3 + H, O, OH) + bath gas&1.7&5.2\\
    \hline
    Zhukov (this work)&13&8 + bath gas&1.6&19\\
    \hline
Gutheil {\it et al.} \cite{Gutheil}&21&8 + bath gas&2.5&67\\
    \hline
    O'Connaire {\it et al.}  \cite{Henri}&23&8 + bath gas&1.8&71\\
    \hline
Konnov \cite{Konnov2000}&29&8 + bath gases&2.3&36\\

    \hline
    Burke {\it et al.} \cite{DryerNew}&22&8 + bath gases&1.8&35\\

    \hline
		\end{tabular}
		\begin{flushleft}
		 $^{*)}$ The number of equations could exceed the number of reactions because of the possible presence of double reactions and of third-body reactions where the activation energy depends on the collisional partner.
	\end{flushleft}
\end{table}
\setkeys{Gin}{draft=false}
\clearpage
\newpage
\newpage
\newpage
\newpage  

\newpage  
\newpage  
\end{document}